\documentclass[superscriptaddress,twocolumn,showpacs,preprintnumbers,amsmath,amssymb]{revtex4}

\usepackage{graphicx}
\usepackage{dcolumn}
\usepackage{bm}

\begin{document}

\title{\boldmath A Measurement of the $D_s^+$ Lifetime}

\author{J.~M.~Link}
\author{P.~M.~Yager}
\affiliation{University of California, Davis, CA 95616}
\author{J.~C.~Anjos}
\author{I.~Bediaga}
\author{C.~Castromonte}
\author{A.~A.~Machado}
\author{J.~Magnin}
\author{A.~Massafferi}
\author{J.~M.~de~Miranda}
\author{I.~M.~Pepe}
\author{E.~Polycarpo}
\author{A.~C.~dos~Reis}
\affiliation{Centro Brasileiro de Pesquisas F\'isicas, Rio de Janeiro,
RJ, Brazil}
% ...


\collaboration{FOCUS Collaboration}
\affiliation{See \url{http://www-focus.fnal.gov/authors.html} 
for additional author information.}

\date{\today}

\begin{abstract}
A high statistics measurement of the $D_s^+$ lifetime from
the Fermilab fixed-target FOCUS photoproduction experiment is
presented.
We describe the analysis of
the two decay modes, $D_s^+\rightarrow \phi(1020)\pi^+$ and
$D_s^+\rightarrow \overline{K}{}^{\ast}(892)^0K^+$, used for the measurement.
The measured lifetime is
$507.4\pm 5.5~(\mathrm{stat.}) \pm 5.1~(\mathrm{syst.})$~fs using
$8961\pm 105$ $D_s^+\rightarrow \phi(1020)\pi^+$
and $4680\pm 90$ $D_s^+\rightarrow \overline{K}{}^{\ast}(892)^0K^+$ decays.
This is a significant improvement over the present world average.
\end{abstract}

\pacs{13.25.Ft, 14.40.Lb, 14.65.Dw}
\maketitle

Experimental measurements of
charm lifetimes have given
guidance towards a theoretical description of
strong interactions at low energy scales
\cite{Reference:bellinibigi,Reference:Bianco}.
For example, the ratio of
the $D_s^+$ and $D^0$ lifetimes 
gives information on the
{\em weak annihilation} contribution to the total decay
\cite{Reference:Bigi1,Reference:Cheung1,Reference:Nussinov,
Reference:Bigi2}.
Not only is this important for
improving our theoretical understanding of strong 
interactions, 
but
the theoretical tools used to calculate lifetimes are the same or
similar to those used in other areas, for example to extract 
$\vert V_{cs}\vert/\vert V_{cd}\vert$ and $\vert V_{cb}\vert$ 
in charm and bottom decays, or
to calculate the $b$-particle 
lifetimes \cite{Reference:Bigi1b,Reference:Bigi3}.

The data used were collected by the FOCUS collaboration
in the 1996--1997 fixed-target run at Fermi National Accelerator
Laboratory. The FOCUS spectrometer is an upgrade of the spectrometer used
in the E687 photoproduction experiment \cite{Reference:e687nim1}.
In the majority of the data used for this measurement (Run Period $B$)
the vertex region consists of four BeO targets and 16 planes of
silicon strip detectors (SSD).
Two of the SSD planes were placed immediately downstream of
the second target, and two immediately downstream of the fourth
(most downstream) target. 
About 20\%\ of the $D_s^+$ decays used in this analysis were
from data taken earlier in the data-taking
run (Run Period $A$) without the four SSD planes in the
target region \cite{Reference:targetssdnim}.
Momentum analysis was done using five
multiwire proportional chambers
and two magnets with opposite polarities. 
Three multicell threshold \v Cerenkov counters were used for
particle identification
\cite{Reference:focuscerenkovnim}.

The $D_s^+\rightarrow \phi(1020)\pi^+$, $\phi(1020)\rightarrow K^-K^+$ and
$D_s^+\rightarrow \overline{K}{}^{\ast}(892)^0K^+$, 
$\overline{K}{}^{\ast}(892)^0\rightarrow K^-\pi^+$
resonant decay modes \footnote{The charge conjugate
mode is implicitly implied unless otherwise stated.}
are used because they have much better
signal-to-background than inclusive $D_s^+\rightarrow K^-K^+\pi^+$
decays.

All $K^-K^+\pi^+$ candidates are tested to see if they form a
vertex with a confidence level greater than 1\%. The candidate
$D_s^+$ momentum vector is then projected to search for a production
vertex with one or more tracks. As many tracks as possible
are included in the production vertex
so long as the vertex confidence level is larger than 1\%.
The production
vertex is required to be within one of the four targets.
To largely eliminate non-charm backgrounds,
the separation $L$ between the production and decay vertices is
required to be larger than $6\sigma_L$ where $\sigma_L$ is the
calculated
error on $L$.

The $\phi(1020)\pi^+$ ($\overline{K}{}^{\ast}(892)^0K^+$) decay mode candidates
are required to have $K^+K^-$ ($K^-\pi^+$) masses
within two sigma of the
nominal $\phi(1020)$ ($\overline{K}^{\ast}(892)^0$) 
mass. Also the magnitude of $\cos\theta^{\ast}$
must be larger than $0.3$ ($0.6$) for the 
$\phi(1020)\pi^+$ ($\overline{K}{}^{\ast}(892)^0K^+$) decay mode candidates,
where $\theta^{\ast}$ is the angle between
the $K^-$ and the $\pi^+$ ($K^+$) in the 
$\phi(1020)$ ($\overline{K}{}^{\ast}(892)^0$) center-of-mass frame.
For the $\overline{K}{}^{\ast}(892)^0K^+$ decay mode, the $K^+K^-$
invariant mass must not be within 
two sigma
of the
$\phi(1020)$ mass to ensure statistically independent samples
in the two decay modes.
Each track in the $K^-K^+\pi^+$ candidate combination must also
satisfy a minimal \v Cerenkov particle identification criteria.
To further reduce the $D^+\rightarrow K^-\pi^+\pi^+$ contamination
by an additional factor of 15
in the $D_s^+\rightarrow \overline{K}{}^{\ast}(892)^0K^+$ candidate sample,
the kaon with the same sign as the pion is required to
have a kaon probability that is larger than 
its pion probability by a factor 
of 20.

\begin{figure}
\centerline{
\includegraphics[width=3.375in]{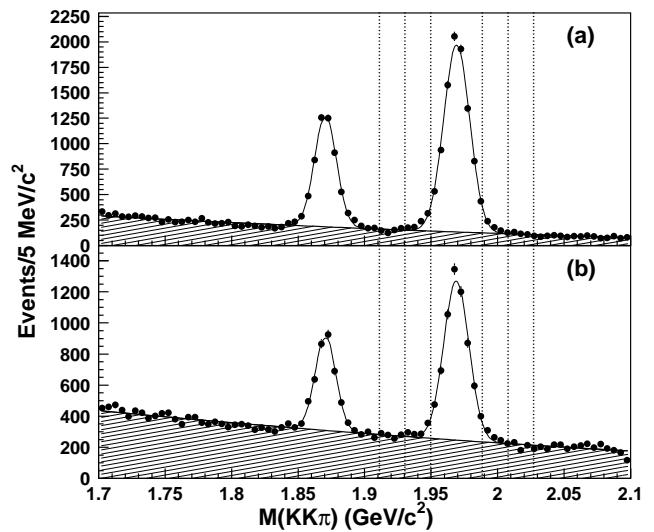}}
\caption{\label{fg_combinedmass} 
The $K^+K^-\pi^+$ invariant mass distributions
for the (a) $\phi(1020)\pi^+$ and (b) $\overline{K}{}^{\ast}(892)^0K^+$ 
decay modes.
The data is given by the points while the line gives the fit with
the hatched region showing the fitted background level.
The vertical dotted lines gives the $D_s^+$ signal and sideband
regions.
}
\end{figure}

The $K^-K^+\pi^+$ invariant mass plots for data are shown in 
Fig.~\ref{fg_combinedmass}. 
The fits shown are to a Gaussian signal and a quadratic background function
which yields $8961\pm 105$ $\phi(1020)\pi$ and
$4860\pm 90$ $\overline{K}^{\ast}(892)^0K$ reconstructed
$D_s^+$ decays. The lifetime analysis uses
$K^-K^+\pi^+$ candidates in a
signal region (SR) that is within
$\pm 2\sigma_m$ of the fitted $D_s^+$ mass, where
$\sigma_m=9.7~\textrm{MeV}/c^2$ is the fitted Gaussian width 
of the $D_s^+$ peak.
Candidates within two symmetric sideband regions (SBR),
each of width $2\sigma_m$
and centered at $\pm 5\sigma_m$ from the fitted $D_s^+$ mass,
are used to represent the lifetime distribution of background in the
SR. The SR and SBR are shown in Fig.~\ref{fg_combinedmass}. 

For the lifetime analysis we use the reduced proper time,
$t^{\prime}=(L-6\sigma_L)/\beta\gamma c$.
The use of the reduced proper time
ensures that only a small acceptance correction to the
lifetime distribution is needed \cite{Reference:Cheung2}. 
The average proper time resolution
for this decay sample (51~fs for Run Period $A$ and
43~fs for Run Period $B$)
is small enough compared to the lifetime to use a binned likelihood
method. We fit the $\phi(1020)\pi$ and $\overline{K}^{\ast}(892)^0K$ 
decay modes
separately.
For each decay mode 
the Run Period $A$ and $B$ data are used in a combined fit.

For each of the run periods $A$ and $B$,
the $t^{\prime}$ distributions for the decays in the SR and SBR
are binned into two separate histograms from 0--3~ps in 60~fs
bins. The observed number of decays in the
{\em i}\,$^\mathrm{th}$ $t^{\prime}$ bin
is $s_i^A$ for the SR and $b_i^A$ for the SBR,
where $A$ refers to Run Period $A$. The
$t^{\prime}$ distribution of the SBR is used as a measure of
the lifetime distribution of background events in the SR. Thus
the expected number of decays ($n_i(A)$) 
in the {\em i}\,$^\mathrm{th}$ $t^{\prime}$ bin of the
SR for Run Period $A$ is given by:
\begin{eqnarray}
n_i(A)=S_A\frac{f_A(t_i^{\prime})e^{-t_i^{\prime}/{\tau}}}
{\sum_i f_A(t_i^{\prime})e^{-t_i^{\prime}/{\tau}}}+
{B_A}{\frac{b_i^A}{\sum_i b_i^A}}\,,
\label{eq:expectednumber}
\end{eqnarray}
and the likelihood for Run Period $A$ is given by
\begin{eqnarray}
{\cal L}_A=\prod_i \frac{n_i(A)^{s_i^A}e^{-n_i(A)}}{s_i^A!}\times
\frac{(\alpha_A B_A)^{N_b^A}e^{-\alpha_A B_A}}{N_b^A!}\,.
\label{eq:likelihood}
\end{eqnarray}
$S_A$ is the total number of signal events, $B_A$ 
is the total number of
background events in the SR, and
$S_A+B_A=\Sigma s_i^A$. The total number of events in the SBR is
$N_b^A=\Sigma_i b_i^A$ and $\alpha_A$ 
is the ratio of the number of events in the
SBR to the number of background events in the SR.
The value of $\alpha_A$ is obtained from the fit to the invariant mass
distribution and is very close to $1$. There is a similar likelihood for
Run Period $B$ and the likelihood that is maximized is
$\cal L = \cal L_A\times \cal L_B$. The fit parameters are
$B_A$, $B_B$, and $\tau$.

\begin{figure}
\centerline{\includegraphics[width=3.375in]{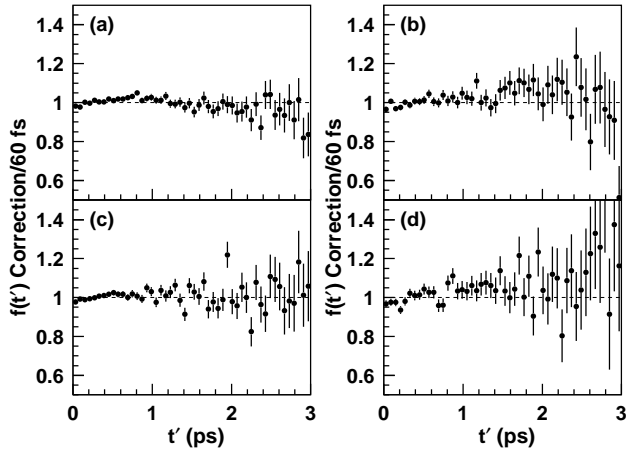}}
\caption{\label{fg_ft} The $f(t^{\prime})$ correction functions for
the two decay modes and two run periods (a) $\phi(1020)\pi$ Run Period $B$;
(b) $\phi(1020)\pi$ Run Period $A$; 
(c) $\overline{K}^{\ast}(892)^0K$ Run Period $B$; and
(d) $\overline{K}^{\ast}(892)^0K$ Run Period $A$.
Deviation from a flat line
indicates the correction from a pure exponential.}
\end{figure}

The effects of geometrical acceptance,
detector, trigger and reconstruction efficiencies,
and absorption are given by the $f(t^{\prime})$
correction function. The $f(t^{\prime})$ 
function is determined using a detailed
Monte Carlo (MC) simulation of the experiment where the production
(using \textsc{Pythia}~\cite{Reference:pythia61})
was tuned so that the production distributions for data and MC matched.
A full coherent $D_s^+\rightarrow K^-K^+\pi^+$ decay was simulated
using results from FOCUS \cite{Reference:E687couplechannel}. 
The $f(t^{\prime})$ distributions are shown in Fig.~\ref{fg_ft}.

\begin{figure}
\centerline{\includegraphics[width=3.375in]{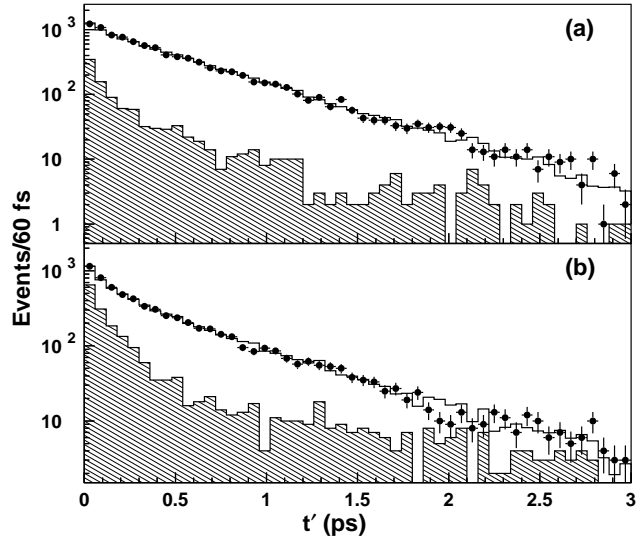}}
\caption{\label{fg_tfit} The
lifetime distribution for all decays in the data SR (points), and
the fit (histogram). The shaded distribution shows the
lifetime distribution of the background component in the SR. The 
data and fit are shown for Run Periods $A$ and $B$ combined in the same
plot for the
two decay modes (a) $\phi(1020)\pi$ and (b) $\overline{K}^{\ast}(892)^0K$.}
\end{figure}

We obtain fitted lifetimes of $507.60\pm 6.46$~fs
and $506.90\pm 10.60$~fs for the $\phi(1020)\pi$ and 
$\overline{K}^{\ast}(892)^0K$ decay
modes, respectively. The fit confidence levels are 2.0\%\ and 0.13\%,
respectively. As discussed below, these low values are expected for
the fitting technique used.
The lifetime
distribution of all decays in the SR are shown in
Fig.~\ref{fg_tfit}\ together with the fit and the level of
background contained in the SR.

Detailed studies were performed to determine the systematic uncertainty in 
the lifetime measurement.

The $f(t^{\prime})$ correction
reduces the 
fitted lifetime by 0.80\%. 
We studied 
the uncertainty in this correction. 
We verified that the MC reproduces the data $D_s^+$ longitudinal and
transverse momenta, the multiplicity of the production vertex, 
$\cos\theta^{\ast}$, and the decay
length and proper time resolutions. A sensitive check of the 
acceptance and efficiency part of the MC
correction was done using high statistics $K_S^0\rightarrow\pi^+\pi^-$ decays.
Short-lived $K_S^0$ decays were reconstructed using the same analysis
methods in the same decay region as the $D_s^+$ decays. Since the
$K_S^0$ lifetime is well known, we can determine the $f(t^{\prime})$ correction
in data and compare it to that obtained in our MC simulation. The agreement
is excellent but was limited in sensitivity
by both data and MC statistics.
Using this sensitivity as the level of the uncertainty in the
$f(t^{\prime})$ correction, we determine a systematic uncertainty due to
this correction of $\pm 0.69$\%.
Possible time dependent systematic effects were looked for by splitting the
data into different time periods and comparing the fitted lifetimes.
We also compared the separate fitted
lifetimes for decays originating from each of the four targets.
No systematic uncertainties were found in these comparisons.

Our limited knowledge of the production and decay of the
$D_s^+$ could contribute to a systematic uncertainty.
This was studied using
different MC simulations where the production parameters and the
resonance substructure of the decay were varied over reasonable ranges.
Production systematics were also studied by splitting the data into
different bins of total and transverse $D_s^+$ momenta,
primary vertex multiplicity, and by comparing the fitted lifetimes for
particles and anti-particles. We found no evidence of systematic
effects, however,
variations of the fitted lifetimes in these studies
led us to assign a systematic uncertainty of
$\pm 0.23$\%\ due to our limited knowledge of $D_s^+$ production
and decay.

The systematic uncertainty due to absorption of the 
$D_s^+$ and daughter particles was determined by varying the
$D_s^+$ interaction cross-section \footnote{The $D_s^+$
cross section is taken to be half the neutron cross section.}
by 100\%\ and the
daughter particle interaction cross-sections by 50\%\ 
in the MC. It was
also studied by comparing the lifetimes of decays occurring inside
and outside of the target, and by comparing the lifetimes for decays
where the $D_s^+$ was produced in the upstream half of each target
with those produced in the downstream half of the same target.
We determined a systematic uncertainty of $\pm 0.38$\% 
due to absorption.

In order to use the reduced proper time we must be able to correctly
model our proper time resolution.
This was verified by comparing the distributions
for data and MC and by studying splits of the data sample 
that can be sensitive
to resolution effects. 
This included splits in $t^{\prime}$ resolution
and reconstructed invariant mass
and varying the $t^{\prime}$ bin width and changing the fitted range
for $t^{\prime}$.

The lifetime distribution of the background in the SR
is assumed to be well represented by the SBR.
The uncertainties that 
arise because of this assumption were determined by a number of
studies.

The contamination from
$D^+\rightarrow K^-\pi^+\pi^+$,
and $\Lambda_c^+\rightarrow K^-p\pi^+$ decays
misidentified as $K^-K^+\pi^+$ decays were determined in our sample.
As a fraction of the total 
$D_s^+\rightarrow \phi(1020)\pi$ 
($D_s^+\rightarrow \overline{K}^{\ast}(892)^0K$)
signal in the SR,
we found the above two backgrounds contribute
0.7\% (3.0\%) and 0.2\%\ (2.5\%), respectively.
The small contribution of
these reflection backgrounds mean they
give rise to insignificant uncertainties. This was verified in a test by
using stronger particle identification and by
explicitly eliminating them by cutting out the appropriate mass regions.

The background lifetime uncertainty was further investigated by using
symmetric sidebands of different widths ($2$--$4\sigma_m$), and located
at different separations from the signal region ($\pm 4$ to
$\pm 6\sigma_m$). 
Variations in particle identification and vertexing selection were used
to study the lifetime with
changes in the signal/background ratio and this showed no
significant effect. 

Finally, an independent analysis which did not rely on
knowledge of the background lifetime distribution was performed. In this
analysis the data were split into fifteen 150~fs wide reduced proper time 
bins from 0--2.25~ps and
one 750~fs wide bin from 2.25--3~ps. 
The numbers of $D_s^+\rightarrow \phi(1020)\pi^+$ and 
$\overline{K}{}^{\ast}(892)^0K^+$
decays in each bin were determined in a mass fit and the yields 
for the two decay modes separately fitted
to an exponential decay distribution modified by a
$f(t^{\prime})$ correction function. 
No significant systematic effects were found.
To account for the variations of the lifetime in the
various background studies we
assign a
background systematic uncertainty of $\pm 0.16$\% ($\pm 1.09$\%)
for the $\phi(1020)\pi$ ($\overline{K}^{\ast}(892)^0K$) decay mode.

A mini-Monte Carlo test of the fit method was done to check the accuracy
of the fit errors and to look for possible biases in the fit.
We generated
1000 independent data samples from the best fit to the data.
Each of these data samples were fit as real data. Distributions of fitted
values showed a small bias in the fitted lifetime which is accounted for 
in a Fit Method
systematic uncertainty (see Table \ref{tb_syst}).
This study also showed
that the fit error
underestimates the true statistical uncertainty 
in the $\phi(1020)\pi$ ($\overline{K}^{\ast}(892)^0K$) mode by 4.8\%\ (14.7\%).
The statistical errors on the lifetimes quoted in this paper 
have been corrected for this effect. These fit properties are due to
neglecting statistical variations in the shape of the background in the
SR, which is fixed in the fit to be equal to the measured background in
the SBR as given by
$b_i^{A,B}$.
This also leads to the small reported fit confidence levels.
When bin-to-bin
fluctuations in the observed $b_i^{A,B}$ are accounted for in the
calculation of the $\chi^2$,
the fit confidence levels for $\phi(1020)\pi$ and $\overline{K}^{\ast}(892)^0K$
become 8.8\%\ and 16\%, respectively.
The goodness of fit is further verified by using 
background lifetime distributions
taken from fits to the data sideband distributions instead of
the actual ($b_i^{A,B}$) entries. This fit method gives 
consistent lifetimes and fit confidence
levels of 9.2\% and 18\%
for $\phi(1020)\pi$ and $\overline{K}^{\ast}(892)^0K$, respectively.

\begin{table}
\caption{\label{tb_syst}Contributions to the systematic uncertainty.}
\begin{ruledtabular}
\begin{tabular}{lccc}
Contribution & \multicolumn{3}{c}{Systematic Uncertainty (\%)} \\
             & $\phi(1020)\pi$  & $\overline{K}^{\ast}(892)^0K$  & Combined \\
\hline
Acceptance  & $\pm 0.69$ & $\pm 0.69$ & $\pm 0.69$\\
Production  & $\pm 0.23$ & $\pm 0.23$ & $\pm 0.23$\\
Absorption  & $\pm 0.38$ & $\pm 0.38$ & $\pm 0.38$\\
Backgrounds & $\pm 0.16$ & $\pm 1.09$ & $\pm 0.41$\\
Fit Method  & $\pm 0.12$ & $\pm 1.07$ & $\pm 0.38$\\
Time Scale \cite{Reference:e831lclt}
            & $\pm 0.11$ & $\pm 0.11$ & $\pm 0.11$\\
\hline
Total       & $\pm 0.85$ & $\pm 1.74$ & $\pm 1.00$ \\
\end{tabular}
\end{ruledtabular}
\end{table}

\begin{table}
\caption{\label{tb_compare}Comparison of 
$D_s^+$ lifetime measurements and $\tau(D_s^+)/\tau(D^0)$ ratios.
The PDG value of the $D^0$ lifetime \cite{Reference:pdg2004}
is used in the ratios, except for the FOCUS ratio for which the
FOCUS measurement is used \cite{Reference:focusd0lifetime}.
}
\begin{ruledtabular}
\begin{tabular}{lcc}
Experiment & $\tau(D_s^+)$~fs & $\tau(D_s^+)/\tau(D^0)$ \\
\hline
E687 \cite{Reference:e687dslt}     & $475\pm 20\pm 7$ & $1.158\pm 0.052$\\
E791  \cite{Reference:e791dslt}    & $518\pm 14\pm 7$ & $1.262\pm 0.038$\\
CLEO II.5 \cite{Reference:cleodslt}& 
                       $486.3\pm 15.0^{+4.9}_{-5.1}$  & $1.185\pm 0.039$\\
SELEX \cite{Reference:selexdslt}   & $472.5\pm 17.2\pm 6.6$& $1.152\pm 0.045$\\
FOCUS (this result)                & $507.4\pm 5.5\pm 5.1$ & $1.239\pm 0.017$\\
\end{tabular}
\end{ruledtabular}
\end{table}

We have measured the $D_s^+$ lifetime to be
$507.4\pm 5.5~(\mathrm{stat.}) \pm 5.1~(\mathrm{syst.})$~fs using 
$8961\pm 105$ $D_s^+\rightarrow \phi(1020)\pi^+$
and $4680\pm 90$ $D_s^+\rightarrow \overline{K}{}^{\ast}(892)^0K^+$ decays
from the Fermilab FOCUS
photoproduction experiment.
We also determined
the ratio of our $D_s^+$ lifetime to
our $D^0$ lifetime \cite{Reference:focusd0lifetime}
to be
$1.239\pm 0.014~(\mathrm{stat.}) \pm 0.009~(\mathrm{syst.})$,
where the systematic uncertainty in the ratio is reduced by eliminating
common systematic contributions.
Table~\ref{tb_compare}\ compares our measurement with 
recent published results.
This measurement represents a significant
improvement in precision.

We wish to acknowledge the assistance of the staffs of Fermi National
Accelerator Laboratory, the INFN of Italy, and the physics departments of the
collaborating institutions. This research was supported in part by the U.~S.
National Science Foundation, the U.~S. Department of Energy, the Italian
Istituto Nazionale di Fisica Nucleare 
and Ministero dell'Istruzione, dell'Universit\`a e della Ricerca,
the Brazilian Conselho Nacional de Desenvolvimento
Cient\'{\i}fico e Tecnol\'ogico, CONACyT-M\'exico,
the Korean Ministry of Education, and the Korean Science and
Engineering Foundation.


\end{document}